\documentstyle[epsfig,amssymb,twocolumn]{mn2e}
\newif\ifAMStwofonts
\title[Thickness of discs in MOND]
{The thickness of H\,{\sc i} in galactic discs under 
MOND: theory and application to the Galaxy}
\author[S\'anchez-Salcedo et al.]
{F.~J.~S\'anchez-Salcedo$^{1}$\thanks{E-mail:jsanchez@astroscu.unam.mx},
K.~Saha$^{2,3}$ and C.~A.~Narayan$^{4}$\\
$^{1}$Instituto de Astronom\'{\i}a, Universidad Nacional 
Aut\'onoma de M\'exico, Apt.~Postal 70
264, C.P. 04510, Mexico City, Mexico\\
$^{2}$Department of Physics, Indian Institute of Science, Bangalore 560012,
India\\
$^{3}$Raman Research Institute, Bangalore 560080, India\\
$^{4}$Astronomisches Institut, Ruhr-Universitaet Bochum, Universitaetsstrasse
150, 44780 Bochum, Germany}

\begin{document}

\date{Accepted xxxx Month xx. Received xxxx Month xx; in original form
2007 July 20} 
\pagerange{\pageref{firstpage}--\pageref{lastpage}} \pubyear{2007}
\maketitle

\label{firstpage}

\begin{abstract}
The outskirts of galaxies are a very good laboratory 
for testing the nature of the gravitational field at low accelerations.
By assuming that the neutral hydrogen gas is in hydrostatic equilibrium in 
the gravitational potential of the host galaxy, the observed 
flaring of the gas layer can be used to test modified gravities. 
For the first time we construct a simple framework to derive
the scaleheight of the neutral hydrogen gas disc in the MOND scenario 
and apply this to the Milky Way. 
It is shown that using a constant gas velocity dispersion of 
$\sim 9$ km s$^{-1}$, MOND   
is able to give a very good fit to the observed H\,{\sc i} flaring 
beyond a galactocentric distance of $17$ kpc up to the last measured
point ($\sim 40$ kpc). Between $10$ and $16$ kpc,
however, the observed scaleheight is about $40\%$ more than
what MOND predicts for the standard interpolating function and
$70\%$ for the form suggested by Famaey \& Binney.
Given the uncertainties in the non-thermal pressure support
by cosmic rays and magnetic fields,
MOND seems to be a plausible alternative to dark matter in 
explaining the Milky Way flaring.
Studying the flaring of extended H\,{\sc i} discs in
external edge-on galaxies may be a promising approach to assess 
the viability of MOND.


\end{abstract}


\section{Introduction}
The success of MOdified Newtonian Dynamics (MOND) 
in reproducing the observed velocity rotation 
curves of spiral galaxies without dark matter haloes is amazing. 
Only in about $10\%$ of the roughly $100$
galaxies considered in the context of MOND does the predicted rotation
curve differ significantly from that observed
(e.g., Sanders \& McGaugh 2002). 
Bekenstein (2004) has suggested a theory that leave a door open
to embed MOND within a physically acceptable dynamical framework.
However, other aspects of MOND
in spiral and elliptical galaxies are yet to be investigated.
It would be desirable to see if MOND is able to reproduce the full
three-dimensional structure of galaxies (e.g., Kuijken \& Gilmore
1987; Hernquist \& Quinn 1987; Buote \& Canizares 1994; McGaugh \& 
de Blok 1998; Milgrom 2001; Stubbs \& Garg 2005; Nipoti et al.~2007).

The flaring of the H\,{\sc i} gas layers has been used to
determine the flattenings of dark haloes for a few external
galaxies because it is sensitive to the shape of the dark matter halo 
(e.g., Olling 1996; Becquaert \& Combes 1997).
The alternative theories to dark matter 
may also have observable consequences on the vertical 
structure of galactic discs. For instance, studies of the flaring of 
the H\,{\sc i} discs have been able to rule out the 
magnetic alternative to dark matter because the magnetic
pressure would produce extraordinary thick discs
(Cuddeford \&  Binney 1993; S\'anchez-Salcedo \& Reyes-Ruiz 2004). 
In modified gravities such as MOND, where the gravity comes mainly 
from the disc, the potential
may be much flatter leading to far more thin discs (e.g., S\'anchez-Salcedo
\& Lora 2005).
In this paper we suggest the usage of the flaring of H\,{\sc i} discs
to test modified gravities.

The Milky Way is an attractive target to carry out
self-consistent tests because of the excellent flaring data, large radial 
extend and rather well-known distribution of stars and gas. 
In the conventional Newtonian view, Kalberla (2003) and Kalberla
et al.~(2007) have found no easy way to match a
standard {\it flattened dark halo} to the Galactic flaring data.
They discuss the role of support by non-thermal pressures,
non-axisymmetric mass distributions of dark and baryonic matter,
spiral structures and so on, and conclude that, besides a
massive round dark halo, a self-gravitating dark matter disc with
mass $\sim 2\times 10^{11}$ M$_{\odot}$ plus a dark matter ring
at $13<R<18.5$ kpc with mass $\sim 2.5\times 10^{10}$ M$_{\odot}$
are required. This result is especially intriguing given other
estimates for the shape of the Milky Way's halo that suggest that
it could indeed be rather spherical or even prolate
(e.g., Helmi 2004; Johnston et al.~2005; Law et al.~2005; 
Belokurov et al.~2006; Fellhauer et al.~2006).
In addition, the existence of a massive dark matter disc 
is unexpected in the collisionless CDM paradigm. This might
indicate that more physics should be added in the simulations
of structure formation (e.g., baryonic physics) or the lack
of complete modelling of the flaring problem.

It is commonly argued that MOND is a more falsifiable theory than
CDM hypothesis because
it has less adjustable parameters to fit astronomical observations. 
One expects that if it is not easy to explain the flaring in CDM
theory, it should be even more difficult in MOND.
It is interesting to 
see if, given the measured baryon distribution of the Milky Way,
MOND is able to successfully explain the Galactic flaring.

In section \ref{sec:theory}, we discuss the vertical distribution of
self-gravitating mass layers under MOND. In particular,
we give the semi-analytic framework to derive the radial dependence
of the thickness of the H\,{\sc i} layer in axisymmetric galactic discs.
In section \ref{sec:MW}, we apply our method to derive the expected MOND
H\,{\sc i} flaring in the Milky Way.  By comparing the predictions
with recent determinations of the thickness of
the H\,{\sc i} disc in the Galaxy, we are
able to test the ability of MOND to reproduce the H\,{\sc i}
flaring in an extended radial range ($10$ kpc $\leq R\leq 40$ kpc).

\section{Flaring of discs: Theoretical approach}
\label{sec:theory}
\subsection{Background and motivation}
\label{sec:background}
A possible interpretation of the missing mass in spiral galaxies
is that the flatness of the rotation curves reflects a failure
of the Newton law rather than the existence of vast quantities 
of dark matter. 
Milgrom (1983) suggested that for axisymmetrical systems, 
the {\it real} acceleration at
the midplane of the configuration, $\vec{g}$, is related with
the Newtonian acceleration, $\vec{g}_{N}$, by:
\begin{equation}
\mu(|\vec{g}|/a_{0})\vec{g}=\vec{g}_{N},
\label{eq:algMOND}
\end{equation}
where $a_{0}$ is an universal acceleration of the order of $\sim
10^{-8}$ cm s$^{-2}$, and $\mu(x)$ is a monotonic
and continuous function with
the property that $\mu(x)\simeq x$ for $x\ll 1$ (deep MOND regime)
and $\mu(x)\simeq 1$ for $x\gg 1$ (Newtonian regime).
Although some effort has been made to derive the form of the
interpolating function from basic principles (e.g., Bekenstein 2004),
it is arbitrary but once it is fixed from
astronomical observations, it is universal.
A rather general form is
\begin{equation}
\mu(x)=\frac{x}{(1+x^{n})^{1/n}},
\end{equation}
where $n(x)>0$ and obeys that $n'(x)\leq 0$ for $x\ll 1$ and
$n'(x)\geq 0$ for $x\gg 1$. The simplest case is to assume
$n$ a positive real constant. The exponent $n$ determines how
abrupt is the transition between the Newtonian to the MOND regime.
Since for $n>2$ the interpolating
functions do not differ much, Milgrom (1983) 
selected $n=2$ for simplicity. 
Interestingly, the standard ($n=2$) interpolating function 
succeded in fitting the rotation curves of most external galaxies, 
being the stellar mass-to-light
ratio of the disc the only free parameter (e.g., Sanders 1996; 
Sanders \& McGaugh 2002; Sanders 2004).
Recently, Famaey \& Binney (2005, hereafter FB) suggested 
that $n=1$ is better for MOND to fit 
the inner rotation curve of the Basel model of the Milky Way 
(Bissantz et al.~2003). 
For a sample of galaxies having a gradual transition from the Newtonian
limit in the inner regions to the MOND limit in the outer parts,
Famaey et al.~(2007) and Sanders \& Noordermeer (2007) conclude that with
respect to rotation curve fitting alone, there is no reason to prefer
$n=1$ to $n=2$, it is only the plausibility of the relative $M/L$ values
for bulge and disc as well as the generally smaller global $M/L$, that 
support $n=1$. Moreover, the best fit for the sample of Famaey et al.~(2007)
was obtained for $n=1$. From the $\chi^{2}$ statistics of the global fits,
they derived $n\geq 0.85$ at one-sigma uncertainty. 
Since values $n>2$ seem to be incompatible with the measured terminal velocities
interior to the Solar circle and require disc $M/L$ ratios slightly higher
than the predictions of stellar population synthesis models (McGaugh 2005), 
it turns out that values in the range $0.85 \leq n \leq 2$ are allowed by
current observations, with $n=1$ the most favoured value.

The phenomenological formulation of MOND by Eq.~(\ref{eq:algMOND}) cannot
be applied to an arbitrary self-gravitating system. Bekenstein \&
Milgrom (1984) suggested a Lagrangian theory and also a nonlinear
differential equation for the nonrelativistic gravitational potential
produced by a mass density distribution $\rho$:
\begin{equation}
\vec{\nabla}\cdot\left[\mu\left(\frac{|\vec{\nabla}\Phi|}{a_{0}}\right)
\vec{\nabla}\Phi\right]=4\pi G\rho.
\label{eq:poissonMOND}
\end{equation}
Only for very special configurations  (one-dimensional symmetry --spherical, 
cylindrical or plane symmetric systems-- or Kuzmin discs) 
the MOND field is related to the Newtonian field by the algebraic
relation Eq.~(\ref{eq:algMOND}) (Brada \& Milgrom 1995). 
The modified Poisson equation given by
Eq.~(\ref{eq:poissonMOND}) overcomes many of the conceptual maladies
of the algebraic equation but it is very difficult to solve. 

In a MONDian baryonic Universe $\rho$ is the density distribution
of the seen baryonic matter. A Newtonist, who assumes that $\Phi$ solves
the Poisson equation, will deduce a fictitious dark matter density:
\begin{equation}
\rho_{\rm dm}=(4\pi G)^{-1} \nabla^{2}\Phi-\rho.
\end{equation}
As Milgrom (2001) demostrated, MOND predicts for disc galaxies a
distribution of fictitious dark matter that comprises a dark disc
and a rounder halo. The surface density of fictitious dark matter
in the disc is 
\begin{equation}
\Sigma_{\rm dm}=\left(\frac{1}{\mu(g^{+}/a_{0})}-1\right)\Sigma,
\label{eq:fictitiousdisc}
\end{equation}
where $\Sigma$ is the real (baryonic) surface density of the disc
and $g^{+}$ is the total MOND acceleration just outside the disc.
At large radii where $\mu\ll 1$, the `dark disc' dominates, implying
that Newtonian disc-flaring analyses should indicate highly flattened 
oblate dark haloes.  
Although the initial applications of the flaring technique for 
NGC 891 and NGC 4244 implied highly flattened dark haloes, the 
gas layer flaring method does not return systematically highly
flattened haloes (Olling \& Merrifield 2000; Merrifield 2002;
Narayan et al.~2005 for the case of the Milky Way). 
The question that arises is:
are the predictions of MOND compatible with measurements of the flattening 
of galaxies?
Since it is not possible for MOND to mimic prolate dark matter haloes, 
a clear-cut probe of the existence of a prolate halo in an 
isolated galaxy would be difficult to reconcile with MOND.

\subsection{Scaleheight of one-dimensional planar mass layer in deep MOND}
Assuming pure planar symmetry, the only component different from zero
is $g_{z}$. The vertical hydrostatic equilibrium equation is:
\begin{equation}
\frac{dP}{dz}=-\rho \frac{d\Phi}{dz}.
\label{eq:equix}
\end{equation}
Integrating this equation from $0$ to $z$ and exploiting MOND field
equation (Eq.~\ref{eq:poissonMOND}), we find
\begin{equation}
\int_{0}^{z}\frac{dP}{dz'}\,dz'=-(4\pi G)^{-1}
\int_{0}^{z}\frac{d}{dz'}\left(\mu\frac{d\Phi}{dz'}\right)
\frac{d\Phi}{dz'} dz'.
\end{equation}
In the deep MOND, $\mu\simeq a_{0}^{-1}d\Phi/dz$, and hence the
RHS of the above equation can be integrated immediately:
\begin{equation}
P(z)-P(0)=-\frac{1}{6\pi G a_{0}}\left(\frac{d\Phi}{dz}\right)^{3}.
\label{eq:diffpressures}
\end{equation}
The value of the vertical force $d\Phi/dz$ can be inferred again from
the MOND field equation after
integrating it between $\pm z$, $d\Phi/dz=\sqrt{2\pi G a_{0}\Sigma(z)}$,
where $\Sigma(z)$ is the integrated surface density 
$\Sigma(z)=\int_{-z}^{z}\rho\, dz'$.
Substituting this expression into Eq.~(\ref{eq:diffpressures}),
\begin{equation}
P(z)-P(0)=-\frac{1}{3}\left(2\pi G a_{0}\right)^{1/2}\Sigma(z)^{3/2}.
\end{equation}
For an isotropic and isothermal gas with one-dimensional velocity dispersion
$\sigma$, the midplane pressure is $P(0)=\rho_{0}\sigma^{2}$,
where $\rho_{0}$ is the density at $z=0$.
Adopting the boundary condition $P(\infty)=0$, 
we derive the pressure at the midplane:
\begin{equation}
\rho_{0} \sigma^{2}= 
\frac{1}{3}\left(2\pi G a_{0}\right)^{1/2}\Sigma_{\infty}^{3/2},
\end{equation}
with $\Sigma_{\infty}=\Sigma(\infty)$.
When the scaleheight $z_{0}$ is defined as $z_{0}\equiv
\Sigma_{\infty}/(2\rho_{0})$, it satisfies 
\begin{equation}
z_{0}=\frac{3}{2} \frac{\sigma^{2}}{\sqrt{2\pi G a_{0}\Sigma_{\infty}}},
\end{equation}
for our isothermal self-gravitating layer in the deep MOND regime.
The dependence of $z_{0}$ on $\Sigma_{\infty}$ is obviously different than
in the Newtonian case. In order to facilitate comparison
it is convenient to rewrite this
formula in a more familiar way exploiting the relation 
$\mu_{\infty}\equiv \mu(\infty)=a_{0}^{-1}\sqrt{2\pi G a_{0}\Sigma_{\infty}}$,
\begin{equation}
z_{0}=\frac{3}{4} \frac{\mu_{\infty} \sigma^{2}}{\pi G \Sigma_{\infty}}.
\end{equation}
Therefore, the usual recipe of replacing $G\rightarrow G/\mu$ to
obtain the MOND value from the Newtonian expression overestimates the
scaleheight by a factor $4/3$ in this case.

\subsection{Infinite mass sheet in an external gravitational field}
\label{sec:externalfield}
Suppose now that our mass layer is under a constant external 
field $\vec{g}_{\rm ext}=(g_{x},0,0)$ and that the dynamics
is dominated by $g_{x}$, i.e.~$g_{x}\gg g_{z}$, where $g_{z}$ is
the vertical acceleration created by the self-gravity of the layer. 
In that case the modified Poisson equation (Eq.~\ref{eq:poissonMOND}) reads
\begin{equation}
\rho=(4\pi G)^{-1}\frac{d}{dz}\left(\mu
\frac{d\Phi}{dz} \right),
\end{equation}
with $\mu(g/a_{0})$ and 
$g=(g_{x}^{2}+g_{z}^{2})^{1/2}\approx g_{x}$ in an external-dominated 
field. Therefore
$\mu$ is constant (independent of $x$ and $z$), and thus
\begin{equation}
\rho=\frac{\mu}{4\pi G}\frac{d^{2}\Phi}{dz^{2}}.
\end{equation}
Following the same steps as in the previous subsection, the vertical 
scaleheight of the layer in an external dominated field is
\begin{equation}
z_{0}=\frac{\mu \sigma^{2}}{\pi G \Sigma_{\infty}}.
\end{equation} 
Note that $\mu (g_{x}/a_{0})=$constant. When $g_{x}\gg a_{0}$, we have
$\mu\approx 1$ and we recover Spitzer's formula. If, on the opposite case, 
the layer is in the deep MOND regime ($g_{z}\ll g_{x}\ll a_{0}$), 
then $\mu\simeq g_{x}/a_{0}$ and $z_{0}$ is proportional to $g_{x}$. 
This result has interesting consequences in galactic discs. 
Consider now a region of the disc outside the central region
such that $g_{R}\gg g_{z}$. If $g_{R}$ is mostly determined by 
the inner stellar disc and $g_{z}$ by the local self-gravity
of the H\,{\sc i} layer, then $g_{R}$ plays the role of the external field.
According to the ongoing discussion, if the adopted stellar mass-to-light 
ratio is increased and thus its mass, then the H\,{\sc i} layer in that region
becomes thicker.
This result is reminiscent of the external field effect (Milgrom 1995)
and will be relevant in \S \ref{sec:interpretation}.

\subsection{Gravitational potential in axisymmetric thin discs}

The MONDian Poisson equation (\ref{eq:poissonMOND}) 
is difficult to solve even numerically. 
However, it can be simplified when studying the vertical structure of the 
outer parts of axisymmetric galactic discs.
The field equation can be written as:
\begin{equation}
\mu(x)\nabla^{2}\Phi+ \vec{\nabla} \mu (x)\cdot \vec{\nabla}\Phi=4\pi G\rho,
\label{eq:kanak1}
\end{equation}
where $x=|\vec{\nabla}\Phi|/a_{0}$. 
For an axisymmetric configuration 
Equation (\ref{eq:kanak1}) can be expanded in cylindrical coordinates $(R,z)$
as:
\begin{equation}
\mu(x)\nabla^{2}\Phi+ \left(\frac{\partial \Phi}{\partial R}
\frac{\partial \mu}{\partial R}\right) \left(1+\xi\right)
=4\pi G\rho,
\label{eq:axiPoisson}
\end{equation}
where 
\begin{equation}
\xi=\xi_{\Phi}\xi_{\mu}
\end{equation}
with
\begin{equation}
\xi_{\Phi}(R,z)=\left(\frac{\partial \Phi}{\partial z}\right)
\left(\frac{\partial \Phi}{\partial R}\right)^{-1}=
\frac{g_{z}}{g_{R}},
\end{equation}
and
\begin{equation}
\xi_{\mu}(R,z)=\left(\frac{\partial \mu}{\partial z}\right)
\left(\frac{\partial \mu}{\partial R}\right)^{-1}=
\left(\frac{\partial g}{\partial z}\right)
\left(\frac{\partial g}{\partial R}\right)^{-1}.
\end{equation}
Note that in a smooth disc galaxy $g_{R}<0$, 
and $\partial g/\partial R<0$ (see the Appendix), whereas
${\rm sign}(g_{z})=-{\rm sign}(\partial g/\partial z)=-{\rm sign}(z)$,
implying $\xi<0$.
No approximations at all have been made so far, and Eq.~(\ref{eq:axiPoisson}) 
is fully equivalent to the original one for axisymmetric systems.

The factor $\frac{\partial \Phi}{\partial R} \frac{\partial \mu}{\partial R}$
in Eq.~(\ref{eq:axiPoisson}) can be rewritten as:
\begin{equation}
\frac{\partial \Phi}{\partial R} \frac{\partial \mu}{\partial R}=
\frac{d\mu}{dx} \frac{\partial \Phi}{\partial R} 
\frac{\partial x}{\partial R}\simeq \frac{1}{a_{0}}
\frac{d\mu}{dx} \frac{\partial \Phi}{\partial R} 
\frac{d}{dR}\left(\frac{v_{c}^{2}}{R}\right).
\end{equation}
In the last equality we have used that $\xi_{\Phi}\ll 1$ within
H\,{\sc i} galactic discs (see the Appendix), 
so that $x\equiv |\vec{\nabla}\Phi|/a_{0}=
x(R,0)\sqrt{1+\xi_{\Phi}^{2}}\simeq x(R,0)=v_{c}^{2}(R,0)/(Ra_{0})$. 
Using that 
\begin{equation}
\nabla^{2}\Phi= \frac{1}{R}\frac{\partial}{\partial R}
\left(R \frac{\partial \Phi}{\partial R}\right)+\frac{\partial^{2}\Phi}
{\partial z^{2}}=
\frac{1}{R}\frac{\partial v_{c}^{2}}{\partial R}
+\frac{\partial^{2}\Phi} {\partial z^{2}},
\end{equation}
we are lead to the final modified Poisson equation under MOND dynamics:
\begin{eqnarray}
\mu \frac{\partial ^{2}\Phi}{\partial z^{2}}=4\pi G \rho-
\left(1+\xi\right)\frac{v_{c}^{2}}{a_{0}R}\frac{d\mu}{dx}\frac{d}{dR}
\left(\frac{v_{c}^{2}}{R}\right)
-\frac{\mu}{R} \frac{d v_{c}^{2}}{dR},
\label{eq:MONDpoisson}
\end{eqnarray}
where $v_{c}^{2}\equiv v_{c}^{2}(R,0)$ is the circular velocity at the
midplane of the disc.
Or, equivalently,
\begin{eqnarray}
\frac{\partial ^{2}\Phi}{\partial z^{2}}=4\pi \frac{G}{\mu} \rho-
\left(1+\xi\right)L(x)\frac{d}{dR}\left(\frac{v_{c}^{2}}{R}\right)
- \frac{1}{R}\frac{d v_{c}^{2}}{dR},
\label{eq:MONDpoisson2nd}
\end{eqnarray}
where $L(x)=d\ln\mu/d\ln x \geq 0$ is the logarithmic derivative 
of $\mu$ at the same value of the argument.  For a nearly constant 
rotation curve, the third term of the RHS is very small. 
Combining Eq.~(\ref{eq:MONDpoisson2nd})
with the hydrostatic equilibrium equation (Eq.~\ref{eq:equix}), one
can obtain the vertical distribution for an isothermal disc.

\subsection{Interpretation of the modified Poisson equation}
\label{sec:interpretation}

In order to gain some insight, below we compare the 
different terms of Eq.~({\ref{eq:MONDpoisson2nd}) with the corresponding
terms in the Newtonian equation. For an axisymmetric disc, the Newtonian
Poisson equation is 
\begin{eqnarray}
\frac{\partial ^{2}\Phi}{\partial z^{2}}=4\pi G \rho-
\frac{1}{R}\frac{d v_{c}^{2}}{dR}.
\label{eq:Newtonpoisson}
\end{eqnarray}
The equation above has been used vastly to study the vertical equilibrium
in the standard dark matter scenario (e.g., Olling 1995, 1996; Olling
\& Merrifield 2000; Narayan et al.~2005).
Obviously, the Newtonian equation can be recovered from the 
MONDian equation taking the limit $a_{0}\rightarrow 0$.
In fact, in the Newtonian
limit, $\mu\rightarrow 1$, $d\mu/dx\rightarrow 0$, hence $L(x)\rightarrow 0$
and Eq.~(\ref{eq:MONDpoisson2nd}) simplifies to Eq.~(\ref{eq:Newtonpoisson}).

Comparing Eqs (\ref{eq:MONDpoisson2nd}) and (\ref{eq:Newtonpoisson}),
we see that in the MOND case, the surface density
is replaced by the effective surface density $\Sigma/\mu$, as expected 
from our discussion in sections \ref{sec:background} 
(see Eq.~\ref{eq:fictitiousdisc}) and \ref{sec:externalfield}.
The confining effect of this term is inversely proportional to $\mu$.
As anticipated in section \ref{sec:externalfield}, if we increase the
stellar mass of the disc by adopting a bit larger stellar 
mass-to-light ratio, the rotation velocity $v_{c}$ will increase
correspondingly and hence, the effective gravitational constant 
$G/\mu\propto G/v_{c}^{2}$ becomes smaller at these radii. 

Interestingly, Eq.~(\ref{eq:MONDpoisson2nd}) contains a term 
proportional to $L(x)$; we will refer to it as the $L$-term. 
In principle, this term has no analog in Newtonian dynamics.
In order to understand its nature, consider the
region of a certain galaxy where the rotation curve is
flat, i.e.~$v_{c}(R)=$const, and
write Eq.~(\ref{eq:MONDpoisson2nd}) in terms of the vertical 
and angular frequencies:
\begin{equation}
\nu^{2}=4\pi \frac{G}{\mu} \rho+
\left(1+\xi\right)L(x)\Omega^{2},
\label{eq:nuMOND}
\end{equation}
where $\Omega (R)=v_{c}/R$.
The $L$-term becomes important in the vertical equilibrium
configuration when 
\begin{equation}
\left(1+\xi\right)\mu(x)L(x) \Omega^{2}
\gtrsim 4\pi G\rho.
\end{equation}
Centainly, the $L$-term is crucial when the centripetal
acceleration is sufficiently low that we are in the MOND regime,
$\mu(x)\approx x$, and for midplane densities lower than
the critical density given by:
\begin{equation}
\rho_{c}(R)=\left(1+\xi\right)\frac{v_{c}^{2}}{Ra_{0}}
\frac{\Omega^{2}}{4\pi G}.
\end{equation}
Since the asymptotic velocity in MOND is $v_{c}^{4}=GM a_{0}$, where
$M$ is the total mass of the disc, the critical density can be rewritten
as $\rho_{c}(R)=(1+\xi)M/(4\pi R^{3})$.

As a reference number, the critical density for a galaxy with $v_{c}=220$
km s$^{-1}$ is $2.3(1+\xi)\times 10^{-3}$ M$_{\odot}$ pc$^{-3}$ at $R=20$ kpc. 
For Galactic parameters,
we find that $|\xi|\approx 0.03$ at $20$ kpc (using
our estimate, Eq.~(\ref{eq:xiestimate}), with
a HWHM scale $h=500$ pc and $\sigma=9$ km s$^{-1}$).
Since the midplane H\,{\sc i} volume density at $20$ kpc 
is $\sim 0.9\times 10^{-3}$ M$_{\odot}$ pc$^{-3}$,
we expect the $L$-term to be important or even dominant
at the very outer disc, say $R\gtrsim 20 (v_{c}/220 {\rm km\, s^{-1}})$ kpc.
At these radii, $\xi$ can be neglected as compared to $1$ and
 Eq.~(\ref{eq:nuMOND}) can be written as:
\begin{equation}
\nu^{2}=4\pi \frac{G}{\mu} \sum_{i=1}^{3}\rho_{i}+ \Omega^{2},
\label{eq:nuMOND2nd}
\end{equation}
where we have explicitly quoted that the baryonic density 
is the sum of three components, the stellar, atomic (He corrected)
and molecular hydrogen mass densities.
Stubbs \& Garg (2005) have proposed to use jointly the rotation velocity
and velocity dispersion of galactic discs to test the self-consistency
of MOND. From our analysis, we learn that $\mu$ can be derived from
the observed vertical velocity dispersion $\sigma_{z}$,
the scaleheight of the disc $h$ and the surface density $\Sigma$,
according to the following relation 
\begin{equation}
\sigma_{z}^{2}=\pi\frac{G}{\mu}\Sigma h+\frac{1}{2}L(x)\Omega^{2}h^{2}.
\end{equation}
\subsection{L-term and the dark matter halo}

Imagine a Newtonist and a MONDianist trying to explain the dynamics of a particular
galaxy, i.e.~the same $\Phi$. Since they use different field equations, they
infer different mass distributions.
Newtonists add a spherical dark halo of the form 
\begin{equation}
\rho_{h}(r)\approx \frac{v_{c}^{2}}{4\pi G r^{2}}
\end{equation}
to explain the flatness of
the rotation curves in the outer parts of disc galaxies.
In the following we argue that the $L$-term is the MOND counterpart
of the Newtonian spherical dark halo. We start noting that, with this dark halo,
the conventional Poisson equation at large radii reads:
\begin{eqnarray}
\frac{\partial ^{2}\Phi}{\partial z^{2}}=4\pi G 
\left(\sum_{i=1}^{3}\rho_{i}+\rho_{h}\right).
\label{eq:vNewton}
\end{eqnarray}
Substituting for $\rho_{h}$ into Eq.~(\ref{eq:vNewton}) with $v_{c}^{2}=\Omega^{2}r^{2}$
we obtain
\begin{eqnarray}
\frac{\partial ^{2}\Phi}{\partial z^{2}}=4\pi G 
\sum_{i=1}^{3}\rho_{i}+\Omega^{2},
\end{eqnarray}
or written in terms of the frequencies:
\begin{equation}
\nu^{2}=4\pi G \sum_{i=1}^{3}\rho_{i}+ \Omega^{2}.
\label{eq:nuNewton}
\end{equation}
Comparing Eqs (\ref{eq:nuMOND2nd}) and (\ref{eq:nuNewton}) we see that,
in the outer parts of galaxies where the dynamics is
in the deep MOND regime, the $L$-term mimics the vertical pull created by
the spherical dark halo in the traditional Newtonian scenario (see also Milgrom 2001).
In summary, MOND predicts for disc galaxies a fictitious round dark halo
and a fictitious dark disc. The resultant flattened potential along the vertical axis
implies that, if MOND is correct, Newtonists should
only infer {\it oblate} dark matter haloes around disc galaxies. 
The inference of prolate ($q>1$) dark haloes in isolated spiral
galaxies would suffix to rule out `classical' MOND. Of course,
MOND is not fundamentally incompatible with the existence of unseen
matter and one might argue that an unseen prolate spheroid of stars
can account for the Newtonian inference of a prolate dark halo,
but this would weaken its appeal.

\section{Application: The Milky Way}
\label{sec:MW}
One advantage of studying our Galaxy as compared to other external
galaxies is that we know quite well its baryonic mass.
For no other galaxy do we have information of comparable quality.
Using the terminal velocities, Famaey \& Binney (2005) were able to put 
constraints on the form of the interpolating
function given the measured baryon distribution.
In addition, the new All Sky H\,{\sc i} Survey (the Leiden-Argentine-Bonn [LAB] 
survey as it is known; Kalberla et
al.~2005) provides us a reliable three dimensional structure of the
H\,{\sc i} disc of the Milky Way up to a minimum of $40$ kpc 
from the Galactic centre. 
Currently, this is the most sensitive Milky Way H\,{\sc i} line survey
with the most extensive coverage both spatially and kinematically.
Since the stellar contribution drops down rapidly in the outer region,  
this huge H\,{\sc i} disc can
serve as an excellent probe for testing alternative theories to dark matter
and thereby the nature of gravity. In this analysis we use MOND (as opposed
to dark matter) as a possible candidate for reproducing the thickness
of the H\,{\sc i} layer of our Galaxy and hence test its 
standpoint in this regard. 

\subsection{The method}
We have derived the vertical distribution of the H\,{\sc i} layer
following the same simplifying assumptions than in the standard 
procedure implemented in Olling (1995). The system is 
assumed to be in a (perhaps turbulent) steady state
which can be described by a Jeans-like equation.
The velocity dispersion tensor of the gas, which includes
thermal and macroscopic gas motions, is assumed to be round and constant
(isothermal) with $z$ as well as $R$.
The maps of the second moment (or velocity dispersion)
in external galaxies
display a patchy distribution (e.g., Petric \& Rupen 2007) with regions
of relatively low velocity dispersion ($5$ km s$^{-1}$) and 
high velocity dispersion ($12$ km s$^{-1}$). Therefore, the assumption 
of constant velocity dispersion should be reasonable
after averaging over azimuth. 
Under these assumptions, the thickness of the Milky Way's gas
layer is dictated by the hydrostatic balance between the pull of
gravity toward the Galactic plane and the pressure forces acting
on the gas:
\begin{equation}
\sigma_{i}^{2}\frac{\partial \rho_{i}}{\partial z}=
-\rho_{i}\frac{\partial \Phi}{\partial z}.
\label{eq:equilibrium}
\end{equation}
If the interstellar gas is in the form of small dense clumps,
the vertical component of its velocity dispersion should be 
smaller than the radial component, in the same manner than stars.
Even so, the term due to the tilt of the velocity dispersion tensor 
is expected to be very small (Olling 1995; Becquaert \& Combes 1997;
Kalberla 2003).
The gradient of the mean pressure of the hot gas ($T>10^{5}$ K) is
negligible within the thickness of the the H\,{\sc i} disc 
because the scaleheight of hot gas is much larger than the scaleheight
of H\,{\sc i}.
For instance, the decrease of hot gas density away from the plane
is consistent with a local scaleheight of $3.2$ kpc. 
Overpressured bubbles of hot gas formed by supernovae explosions may
be a source of driving porosity in the ISM and turbulent motions in 
the H\,{\sc i}. Porosity may modify the hydrostatic properties of the
gas layer. However, for a turbulent ISM,
the effect of the hot medium pressure is essentially included
in the {\it turbulent} velocity dispersion.

To follow the standard technique as close as possible in order
to facilitate comparison with previous work, the magnetic and cosmic 
ray pressures are neglected. This is probably the most fragile assumption.
To separate variables, we will postpone investigating the effects
of non-thermal pressure terms until section \S \ref{sec:nonthermal}.

Equation (\ref{eq:equilibrium}) must be solved under boundary
conditions at high $z$, which are given by the properties of the
embedding ambient medium.
One can see that reasonable hot halo components in hydrostatic equilibrium,
as that derived by Benjamin \& Danly (1997), have little impact on the
H\,{\sc i} scaleheight.  Not only the pressure by the hot
halo gas but also any ram pressure exerted by the
intergalactic medium onto the disc is also thought to have
an ignorable effect on the vertical distribution of the H\,{\sc i} layer.

To date, most authors have adopted these assumptions to determine
the distribution of the gas (e.g., Olling 1995, 1996; 
Becquaert \& Combes 1997; Olling \& Merrifield 2000; Narayan \& Jog 2002;
Kalberla 2003; Narayan et al.~2005; Abramova \& Zasov 2007). 
Although the limitations of these suppositions are well-documented
in the literature, the reader is referred to \S \ref{sec:limitations} for
a discussion.

\subsection{Mass model}
\label{sec:massmodel}
In order to estimate the thickness of the H\,{\sc i} layer, the circular 
velocity should be known (see Eq.~\ref{eq:MONDpoisson2nd}). 
Given the measured baryon distribution, the circular velocity can be 
determined.
In this sense, MOND has less free parameters than the hypothesis
of dark haloes.
We consider the Galaxy to be composed of a bulge, a stellar 
disc and ISM in the form of H\,{\sc i} and H$_2$ layers.
For our Galaxy they are all based on well-known observations as described below. 
Our analysis treats the gas and the stars as axisymmetric distributions.
Therefore, the resulting quantities should be considered as average values over
azimuth.

\subsubsection{Stellar components}
The density profile of the bulge is assumed to be described by a 
spherical Plummer-Kuzmin model:
\begin{equation}
\rho_{b}(r)=\frac{3M_{b}}{4\pi R_{b}^{3}}\left(1+\frac{r^{2}}{R_{b}^{2}}\right)^{-5/2},
\end{equation}
where the bulge mass $M_b=3.2\times 10^{10}$ M$_\odot$ and the bulge
scalelength $R_b=2.5$ kpc. Inclusion of this simple bulge  
reproduces a reasonable looking rotation curve of the Galaxy. 
The contribution of the bulge to the H\,{\sc i} scaleheight in the 
outer parts ($R>10$ kpc) of the Galaxy is negligible.   
For instance, for a mass $M_b=1.5\times 10^{10}$ M$_\odot$ (e.g.,
Flynn et al.~2006), the scaleheight changes $<4\%$.

The stellar disc is modelled as an exponential surface density 
distribution with a
central surface density $\Sigma_{0}$ and disc scalelength $R_d$ such that:
\begin{equation}
\Sigma_{\star}(R) = \Sigma_{0}e^{-R/R_d},  
\end{equation}
where  $\Sigma_{0}=640.9$ M$_{\odot}$pc$^{-2}$ and $R_d=3.2$
kpc (Mera et al.~1998; L\'opez-Corredoira et al. 2002;
Larsen \& Humphreys 2003; Sommer-Larsen et al. 2003), 
implying a total stellar mass of $4.1\times 10^{10}$
M$_{\odot}$ in the disc. 
The local stellar surface density (at $R =$R$_\odot = 8.5$
kpc) turns out to be 45 M$_{\odot}$ pc$^{-2}$, which when combined with the
corresponding value for gas (H\,{\sc i} + H$_2$) of $7$ M$_{\odot}$ pc$^{-2}$, 
gives the
total local surface density inferred by most observations (e.g., Holmberg \&
Flynn 2004). The vertical velocity dispersion of the stars 
as a function of $R$ is taken from Lewis \& Freeman (1989).
 
\subsubsection{Gaseous components and the flaring of the H\,{\sc i} layer}

The H\,{\sc i} surface density (He corrected\footnote{Our analysis does include 
a $30\%$ per cent contribution by mass from helium.})
as a function of the galactocentric radius 
is taken as an input for the disc model.  
As said before, the H\,{\sc i} data comes from the most recent comprehensive 
H\,{\sc i} survey of our Galaxy, the LAB survey (Kalberla et al.~2005). 
Here we use the average observed flaring (of North and South hemispheres) 
results for a flat rotation curve with $v_{c}=220$ km s$^{-1}$
and $R_{\odot}=8.5$ kpc (the current IAU standard values), 
to compare with the predicted scaleheight in MOND.

The vertical velocity dispersion of H\,{\sc i}, $\sigma_{\rm HI}$, 
outside the solar circle 
is hard to measure, and so not yet known. 
Observations of external galaxies imply that $\sigma_{\rm HI}$
is typically constant along galactocentric radius with a  
value $7$--$9$ km s$^{-1}$ (e.g., Sellwood \& Balbus 1999), 
or declines slightly with radius (Dib et al. 2006, and references therein).
Since in the inner Galaxy the H\,{\sc i} is observed to have a 
velocity dispersion 
of $9.2\pm1$ km s$^{-1}$ independent of radius (Malhotra 1995) and
since there is evidence that the velocity dispersion in the outer 
Galaxy equals the
value inside the solar circle (Blitz \& Spergel 1991), Olling \& Merrifield
(2000, 2001) assumed a constant $\sigma_{\rm HI}$ of $9.2$ km s$^{-1}$
as a more likely value. However, a slightly declining velocity dispersion 
could also be admissible.  
In order to calculate the scaleheight we consider
models across a range of values $7$ km s$^{-1}\leq 
\sigma_{\rm HI}\leq 9$ km s$^{-1}$, constant with radius.

The H$_2$ surface density as a function of $R$
was taken from Wouterloot et al.~(1990) and its vertical velocity dispersion is 
$5$ km s$^{-1}$ (Clemens 1985).

\subsection{MOND parameters and the rotation curve} 
Milgrom's formula (Eq.~\ref{eq:algMOND}) has been used extensively to fit the 
rotation curves of external galaxies. 
The MOND circular speed $v_{c}$ in the disc mid-plane is related 
to its Newtonian counterpart (without any dark matter), $v_{c,N}$, 
by the following equation:
\begin{equation}
\mu(x)v_{c}^2  \:=\: v_{c,N}^2 
\end{equation}
(Brada \& Milgrom 1995) where $x=v_{c}^{2}/(Ra_{0})$ and the Newtonian rotation speed in the 
mid-plane of the Galaxy is given by:
\begin{equation}
v_{c,N}^2  \:=\: v_{\rm disc}^2 + v_{\rm bulge}^2 + v_{\rm gas}^{2}. 
\end{equation}
For a set of external galaxies with high quality rotation curves, Begeman et
al.~(1991) derived $a_0 = (1.2 \pm 0.27) \times 10^{-8}$ cm s$^{-2}$. This has
been the preferred value for MOND studies that followed (Sanders \&
Verheijen 1998; Famaey \& Binney 2005). However, Bottema et al.~(2002) argue
based on Cepheid-based distance scale to UMa cluster of galaxies that the
value of $a_0$ should be adjusted to $0.9 \times 10^{-8}$ cm s$^{-2}$. 
Adopting the mass model described in \S \ref{sec:massmodel} and
the relation between the Galactic constants,
$\Theta_{0}/R_{\odot}=(27\pm 2.5)$ km s$^{-1}$kpc$^{-1}$
(Kerr \& Lynden-Bell 1986; Reid et al. 1999), reasonable
rotation curves are generated for $a_0$ in the range
of $0.8\times 10^{-8}$ to $1.2 \times 10^{-8}$ cm s$^{-2}$. 
Figure \ref{fig:rc} shows the rotation curves for the interpolating
functions with $n=1$ and $n=2$ and different $a_0$.  
The radial upper
limit of 40 kpc is set by the reliability of H\,{\sc i} data.  
At the solar point R$_{\odot}=8.5$ kpc,
the circular speed lies in the range of 215 to 243 km s$^{-1}$, conforming with
the accepted range of ratio of Galactic constants. This implies that the model
parameters assumed above are appropriate. 
Note that the differences in the rotation curves are small
at large radii because the asymptotic value of the  
rotation speed is given by $v_{c}(\infty)=(G M a_0)^{1/4}$ 
for both interpolating functions.  

We like to point out that the assumed rotation curves behind the 
theoretical and observed scaleheight curves do not exactly match. In the
case of  H\,{\sc i} observations, the conversion of brightness 
temperature $T_{B}$, 
to volume density depends on the assumed distance to the Galactic
centre $R_{\odot}$, the local rotation velocity $v_{\odot}$
and the outer rotation curve.  The conventional way 
to go with LAB survey analysis is to assume a flat rotation curve of $220$ 
km s$^{-1}$ with $R_{\odot}=8.5$ kpc. The scaleheight data used in this 
paper is based on these values, whereas 
the rotation curves used to estimate the theoretical flaring are based 
on the adopted mass model of our Galaxy. Ideally, we should use rotation 
curves in Fig.~\ref{fig:rc} to analyse the LAB data.
However, the difference between the two is only $10\%-15\%$
and we will see that, for our purposes here, the   
results are stable against such modest adjustments.
So its effect on the scaleheight will
not compromise our conclusions.

\begin{figure}
{\rotatebox{270}{\resizebox{9cm}{8cm}{\includegraphics{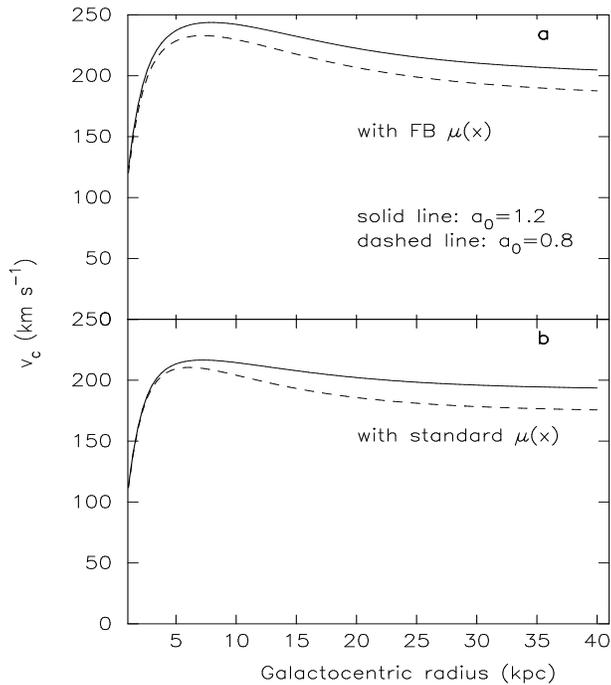}}}}
\caption{Rotation velocity curves for different $a_{0}$, in units
of $10^{-8}$ cm s$^{-2}$, and for the
FB ({\it panel a}) and standard ({\it panel b}) interpolating functions. 
Note that the circular speed $v_c$ is slightly higher for the FB
interpolating function.
MOND produces a falling  rotation curve in the region of interest
($R>10$ kpc).} 
\label{fig:rc}
\end{figure}

\subsection{H\,{\sc i} scaleheight in the Galaxy with MOND}
Using Runge-Kutta method, we solve equations 
(\ref{eq:MONDpoisson2nd}) and (\ref{eq:equilibrium}) numerically
as an initial value problem to obtain 
our object of interest: density as a function
of $z$ at different $R$. 
Readers are referred to Narayan \& Jog (2002) for a
detailed numerical procedure. 
\begin{figure}
{\rotatebox{270}{\resizebox{8cm}{8cm}{\includegraphics{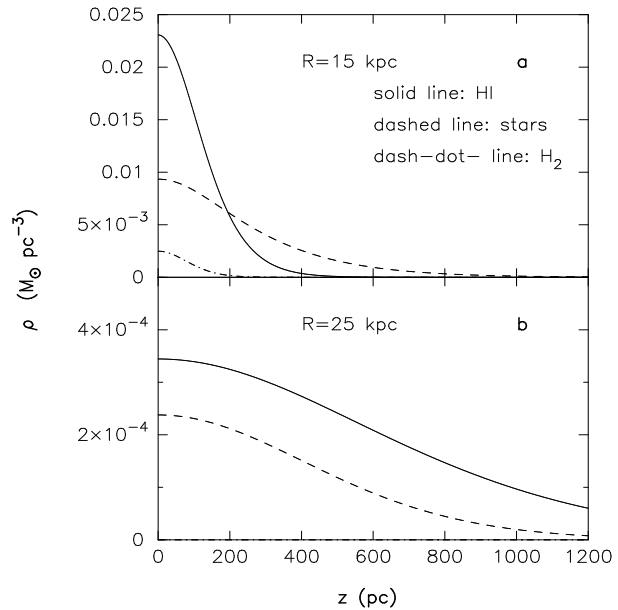}}}}
\caption{Density distribution along the vertical direction for 
H\,{\sc i}, stars and H$_2$ in the outer parts of the Galactic disc
according to MOND for $\sigma_{\rm HI}=7$ km s$^{-1}$, 
$a_{0}=1.2\times 10^{-8}$ cm s$^{-2}$ and $n=1$.}
\label{fig:density}
\end{figure}
As discussed in \S \ref{sec:interpretation}, we can neglect $\xi$ as 
compared to $1$ in Eq.~(\ref{eq:MONDpoisson2nd}) in the whole range of
interest ($10$ kpc $\leq R\leq 40$ kpc) because at large radii where the
$L$-term is important, $\xi \lesssim 0.04$ (see the Appendix). 
In this way we obtain the simultaneous vertical density distributions of stars,
H\,{\sc i} and H$_2$.  

The vertical density distributions of stars and H\,{\sc i} for 
$\sigma_{\rm HI}=7$ km s$^{-1}$, $n=1$ and $a_{0}=1.2\times 10^{-8}$
cm s$^{-2}$ are shown in Fig.~\ref{fig:density}.
Even though we have three disc
components (stellar disc, H\,{\sc i} and H$_2$ layers) having 
different vertical velocity dispersions and
gravitationally coupled under MOND, the vertical density profiles are still
very close to a sech$^2$-function.
As expected, the H$_2$ layer is the most compact one 
followed by the H\,{\sc i} layer and the stellar disc, respectively 
(see Fig.~\ref{fig:density}, upper panel). The
scene is different at large distances from the Galactic centre. The
azimuthally averaged H$_2$ surface density is so low there that it becomes
incorrect to assume it as a continuous layer. It may exist as rare clouds and
so disappear from our picture. 

By solving the equations at regular $R$ intervals (between $10$--$40$ kpc), 
the scaleheight $h$, defined as the HWHM of the vertical 
 H\,{\sc i} density distribution\footnote{Note that the definitions
of z$_0$ and $h$ are different. In this section, it is more convenient 
to use $h$.},
can be plotted as a function of $R$ (Fig.~\ref{fig:scaleheight}). 
The H\,{\sc i} layer undergoes flaring due to a
combination of decreasing total surface density and a steady velocity
dispersion of $7$ km s$^{-1}$, and therefore extends vertically beyond the
stellar disc.  The $L$-term in Eq.~(\ref{eq:MONDpoisson2nd})
becomes similar to the local self-gravity term, $4\pi \mu^{-1}G\rho$,
at $\sim 27$ kpc.  At $R=40$ kpc it turns out to be $4$ times larger than
the local self-gravity when the FB interpolating function
and $a_{0}=0.8\times 10^{-8}$ cm s$^{-2}$ are used. 
While the differences in the flaring between the standard
and the FB interpolating functions are 
not significant beyond $R=27$ kpc, the standard interpolating
function predicts a thicker H\,{\sc i} disc at $10$ kpc $<R<20$ kpc 
(see \S \ref{sec:form} below).  
For $n=2$, $a_{0}=1.2\times 10^{-8}$ cm s$^{-2}$ and $\sigma_{\rm HI}=7$ 
km s$^{-1}$, MOND 
predicts a scaleheight of $\sim 375$ pc at $R=20$ kpc
and $\sim 1625$ pc at $R=40$ kpc.

Now we make a one-to-one comparison between the predicted and the observed 
scaleheights of H\,{\sc i} from the LAB survey. This comparison is possible 
only for the H\,{\sc i} layer because the number counts of both stars
and molecular hydrogen fizzle out at such large radii. 
The circles in Figs.~\ref{fig:scaleheight} and \ref{fig:happy} 
represent the H\,{\sc i} scaleheight 
from the LAB survey data as derived in the reference model 
(see \S 3.2.1 in Kalberla et al.~2007).
For the adopted $\sigma_{\rm HI}$, the predicted
MOND scaleheight always falls below the
LAB values in the entire range considered. Close to $10$ kpc, the 
predicted value is almost $60\%$ below the observed value and at $40$ kpc 
it is about $30\%$ less. The lower $a_0$ slightly 
lifts the curve up but the discrepancy at the interval $10$ kpc$<R<17$ kpc
and beyond $30$ kpc still remains. 

The velocity dispersion in our Galaxy is, of course, uncertain.
In the inner Galaxy, the H\,{\sc i} is observed to have a velocity
dispersion of $9.2\pm 1$ km s$^{-1}$, independent of radius (Malhotra 1995).
If we assume that this value characterizes the turbulent motions of the
gas throughout the Galaxy, the flaring of the H\,{\sc i} layer can be
reproduced well at $R\gtrsim 17$ kpc (see Fig.~\ref{fig:happy}). 
At $25$ kpc $\leq R\leq 32$ kpc, the predicted flaring curve moves away from the
observed one by less than $15\%$. Due to the fact that the flaring in the
northern hemisphere is $\sim 80 \%$ larger than in the sourthern hemisphere,
the mentioned discrepancy of $15\%$ can be attributed to our axisymmetric
assumption. It is important to note that the $L$-term and the local
self-gravity term vary having a different radial dependence and should work out 
in phase to explain the flaring curve. If the $L$-term were not included,
MOND would overestimate the thickness of the H\,{\sc i} disc 
beyond $\sim 25$ kpc. In particular, it would be overestimated by a factor
of $5$ at $R=40$ kpc. The success of MOND in predicting the H\,{\sc i} flaring
at the outer parts is noteworthy because at such large radii the
predictions are very robust to small adjustments on the Galactic mass model,
Galactic constants, the value of $a_{0}$ or the index $n$ of the
interpolating function (see, for instance, Fig.~\ref{fig:scaleheight}). 
The thickness is essentially determined by
the adopted $\sigma_{\rm HI}$. 
If a H\,{\sc i} velocity dispersion of $\approx 9$ km s$^{-1}$ is taken 
at face value, the observed H\,{\sc i} data 
up to $40$ kpc seems to favour the lowest acceptable values  
of the universal acceleration $a_0 \sim 0.8 \times 10^{-8}$ cm s$^{-2}$.

In Fig.~\ref{fig:happy} we see that the observed thickness at 
$10$ kpc $\lesssim R \lesssim 15$ kpc
is still a factor of $1.7$ larger as compared to the MOND prediction.
This factor cannot be attributed to the approximations we made in 
the derivation of Eq.~(\ref{eq:MONDpoisson2nd}).
It is also unlikely that systematic deviations from axisymmetry can account for
that difference because the
average thickness over $0^{\circ}<\phi <180^{\circ}$
is similar to the average value over $180^{\circ}<\phi< 360^{\circ}$
at the interval $10$ kpc $\leq R\leq 15$ kpc (Levine et al.~2006a; 
Kalberla et al.~2007).

It should be borne in mind that the flaring curve and the H\,{\sc i} 
volume density as inferred from the LAB data
are dependent on the Galactic constants and the adopted rotation curve 
and we pick up the $v_{c}=220$ km s$^{-1}$ reference model. Nevertheless,
we established, after some investigation, that the above mentioned 
discrepancy between the predictions and observations
cannot be smeared out by slightly changing the values of $R_{\odot}$
and $v_{\odot}$, or by adopting the rotation curves of Fig.~\ref{fig:rc}
to analyse the LAB data.
In order to sort through a large parameter space, we look at
other parameters that do not appear to be strongly constrained by the data; 
this will be done in the next section. 

\subsection{Effects of incomplete modelling and other uncertainties}
\label{sec:limitations}
In the previous section it was found that in the MOND formulation
the observed disc material gives rise to too thin a thickness
at $10$ kpc $\leq R\leq 15$ kpc. In order to determine the implications
of this result, we need to consider effects of incomplete modelling
and other uncertainties.

\subsubsection{The stellar mass-to-light ratio}
As already said, the $L$-term is small as compared to the self-gravity 
term in the ``inner'' disc ($R\leq 20$ kpc), implying that 
$h\propto \mu/ \Sigma$. 
It is possible to have a thicker H\,{\sc i} disc
in the inner disc if $\Sigma$
is decreased simply by decreasing the stellar mass-to-light
ratio\footnote{The associated 
error in the local 
column density of stars is $\sim 10$ M$_{\odot}$ pc$^{-2}$ (Flynn \& Fuchs
1994; Holmberg \& Flynn 2004). In addition,
strong departures from axisymmetry could mean that the local
surface density may not be representative of the azimuthally
averaged value.} $\Upsilon$. In fact, if $\Upsilon$ is decreased,
the circular velocity and thus $\mu$ lower but the drop in $\mu$ is not
enough to compensate the reduction on $\Sigma$. 
In particular, if we take the FB interpolating function 
with $a_{0}=0.8\times 10^{-8}$ cm s$^{-2}$ and reduce $\Upsilon$ 
to a value such as that $v_{\odot}=200$ km s$^{-1}$, which is quite low
but still compatible with the terminal velocity data (Famaey \& Binney 2005),
$h$ increases about $27\%$ at $R=15$ kpc and about $10\%$ at $R\gtrsim 35$ kpc.
Therefore, the fit is globally better in the latter case but the overcorrection
is not eliminated.

\subsubsection{Non-thermal pressure terms}
\label{sec:nonthermal}

A velocity dispersion $\sigma_{\rm HI}$ of $\sim 9$ km s$^{-1}$ may 
be attributed to the gas random motions.
Potentially, there may be other forces helping to support the H\,{\sc i}
layer as cosmic rays and magnetic fields, which lead to
a larger {\it effective} velocity dispersion $\sigma_{\rm HI}'$
(e.g., Parker 1966; Spitzer 1978; Boulares \& Cox 1990). 
The variation with radius in the pressure support
from magnetic fields and cosmic rays is essentially unconstrained.
The simplest assumption is that, like the turbulent pressure, these
two terms do not vary with radius and so the effective velocity
dispersion is a constant.
Following Spitzer (1978), denote by $\alpha_{B}$ the ratio
between the magnetic pressure arising from the regular field
$P_{B}$ and the kinetic pressure $P_{g}$. 
Similarly, we introduce
$\alpha_{b}\equiv P_{b}/P_{g}$
and $\alpha_{CR}\equiv P_{CR}/P_{g}$, where $P_{b}$ is the pressure
arising from the random field and $P_{CR}$ the pressure due to cosmic rays. 
The scaleheights of the non-thermal pressure terms are $\kappa_{B}$,
$\kappa_{b}$ and $\kappa_{CR}$ times larger than the gaseous scaleheight. 
With this notation, the relative increase in the velocity dispersion is
\begin{equation}
\frac{\sigma_{\rm HI}'}{\sigma_{\rm HI}}=
\sqrt{1+\frac{\alpha_{B}}{\kappa_{B}}+\frac{\alpha_{b}}{\kappa_{b}}
+\frac{\alpha_{CR}}{\kappa_{CR}}}
\label{eq:effectivedisp}
\end{equation}
(e.g., Olling \& Merrifield 2001).
Models and observations of the ISM in the solar vicinity 
and external galaxies are compatible
with approximate equipartition between the energy density of the turbulent
magnetic field and the turbulent energy density of the gas (e.g.,
Zweibel \& McKee 1995;
Beck et al.~1996; Beck 2001; Fletcher \& Shukurov 2001; Beck 2007), hence 
$\alpha_{b}\simeq 0.3$--$0.5$ (see also Kalberla \& Kerp 1998).
Observational constraints on the magnetic field near the Sun
yield $1\lesssim b/B\lesssim 3$ (e.g., Fletcher \& Shukurov 2001 and references therein),
thus $\alpha_{B}=0.05$--$0.5$.  Equipartition arguments suggest
the cosmic ray pressure to be equal to the total magnetic pressure, implying
$\alpha_{CR}\simeq 0.5$--$0.7$. 
The vertical distribution of the non-thermal
components are even more difficult to obtain than is that of the gas components
of the ISM. If the turbulent magnetic field has the same origin
than the turbulent motions, $\kappa_{b}=1$. For the regular magnetic
atmosphere $\kappa_{B}=5$--$10$ are reasonable values. 
The possible values of $\kappa_{CR}$ can be derived by noting 
that the cosmic ray pressure is equal to the total magnetic pressure.

Using Eq.~(\ref{eq:effectivedisp}) with the above range of parameters 
we obtain $\sigma_{\rm HI}'/\sigma_{\rm HI}=1.2$--$1.4$.
This is in accordance to the results of Kalberla \& Kerp (1998) who conclude
that $\sigma_{\rm HI}'/\sigma_{\rm HI} \approx 1.15$  
due to the turbulent magnetic field component, when the pressure contribution
from cosmic rays is not included. 
Ratios $\sigma_{\rm HI}'/\sigma_{\rm HI}\approx 1.4$ are commonly used in
the literature (e.g., Sellwood \& Balbus 1999; Elmegreen \& Hunter 2000).

Since the discrepancy in the scaleheight is $25\%$--$70\%$, depending on the
adopted values of $n$ and $\Upsilon$, values of 
$\sigma'_{\rm HI}/\sigma_{\rm HI}=1.12$--$1.3$ are required to 
account for the thickness of the disc at $10$ kpc $\leq R\leq 15$ kpc.
We conclude that, given the uncertainties, simplified estimates 
indicate that non-thermal pressure support could be sufficient.
Beyond the optical radius $\sim 15$ kpc, where star formation is 
almost nonexistent, energy input into cosmic rays and magnetic fields
from stellar sources is likely to be unimportant. 
Hence, a decreasing $\sigma_{\rm HI}'$ with $R$ is expected.

\subsubsection{Form of the interpolating function}
\label{sec:form}

In section \ref{sec:background}, it was said that 
analyses of the rotation curves of our Galaxy and other external galaxies 
constrain the exponent $n$ of the interpolating function 
in the range $0.85\leq n \leq 2$.  Whilst at the far outer disc
$R>27$ kpc the thickness of the H\,{\sc i} disc remains almost
unchanged when $n$ is varied between $0.85$ and $2$,
the flaring of the disc in the inner disc ($10$ kpc $<R<20$ kpc)
scales approximately as $h\propto \mu$ and so that it depends 
on the adopted value of $n$.
Once the baryonic mass distribution is known,
the ratio between scaleheights derived with $n=2$ and $n=1$
in the inner disc is:
\begin{equation}
\frac{h(n=2)}{h(n=1)}=\sqrt{2}\frac{\left(\sqrt{1+4y^{2}}-1\right)^{1/2}}
{\sqrt{1+4y}-1},
\end{equation}
where $y=(v_{c,N}^{2}/Ra_{0})^{-1}$. In the range of interest
we have $y\gtrsim 1$ implying that 
the scaleheight is about $25\%$ larger when the standard
interpolating function is used than in the FB case. 
Consequently, MOND requires less non-thermal support, 
$\sigma_{\rm HI}'/\sigma_{\rm HI}\simeq 1.18$ for our
standard set of parameters described in \S \ref{sec:massmodel}, to explain
the observed flaring if the standard interpolating function ($n=2$) 
is adopted.

\subsubsection{Outer spiral arms and warps}
It is possible that the H\,{\sc i} layer is slightly out of hydrostatic
equilibrium because of the warp and the presence
of outer spiral arms. The scattering of stars and
compact clouds by the spiral arms hardly increases their vertical velocity
dispersion (e.g., Jenkins \& Binney 1990). However,
if the H\,{\sc i} layer behaves as a smooth disc of gas, 
it develops a combination of a shock and a hydraulic
jump; the gas may shoot up to higher $z$ (e.g., G\'omez \& Cox 2002, 2004). 
Nevertheless, the arms outside the corotation radius are likely weak
and these vertical complications are expected to occur along a 
thin (a few hundred parsecs) region downstream the shock/jump
of the arm.  The observed correlation between the maps of
H\,{\sc i} thickness and surface density --regions of higher 
surface density having a more reduced thickness-- 
(Levine et al.~2006b),
suggests that hydrostatic equilibrium is a good approximation.
By averaging over azimuth, the alteration of the vertical
structure by this jump/shock should not be significant.

It is well-known that the H\,{\sc i} disc of the Galaxy is warped.
A warp is essentially a vertical $m=1$ mode that is maintained
by some external driving force. The S-shape of the warp does not
affect appreciably the local self-gravity in the disc because the
wavenumber of the warp $k\ll 2\pi/h$. 
Since the orbit of a parcel of the disc is inclined to its galaxy's
equatorial plane, its height $z$ above this plane oscillates around
$z=0$ with amplitude $h_{w}(R)$.
It is important to know at what extent this oscillatory
motion may generate vertical compressions or decompressions and
produce an azimuthal dependence on the thickness of the H\,{\sc i} disc.
First, compressions should induce a pressure gradient in the vertical
direction of the disc (probably a shock wave or compression front)
and, consequently, a characteristic asymmetric vertical morphology
is expected (S\'anchez-Salcedo 2004). By contrast, the observed fairly 
symmetric distribution
of the Galactic warp suggests that the gas layer moves vertically
in a solid-like fashion, as if it were incompressible.
This solid-like behaviour is expected
either if the bending modes are long-lived eigen-modes or
they are excited by gravitational torques\footnote{Gravitational 
torques may be produced by
a companion galaxy (e.g., Weinberg \& Blitz 2006)
or by cosmic infall accretion of protogalactic material
in the outer halo (Ostriker \& Binney 1989).}. 
Just for illustration, let us consider the tidal
excitation as the mechanism responsible for maintaining the warp
with amplitude $h_{w}(R)$.
The gravitational perturber
(e.g., the Magellanic Clouds) may exert a tidal stretching on the disc,
but this effect is very small across the thickness of the H\,{\sc i} disc
and hence we only need to study the restoring gravitational force that pulls
the gas back to the disc midplane.
For low amplitudes of the warp $h_{w}/R\ll 1$,
the underlying gravitational restoring
force due to the disc is described by an harmonic potential,
$\Phi(z)=\nu^{2}z^{2}/2$ with $\nu$ the vertical frequency derived in
\S \ref{sec:interpretation}.
Under this potential, one can show that
the oscillatory mode in the vertical direction occurs at
constant thickness. In fact, denoting by $\rho_{eq}(z)$ the hydrostatic
density profile in the absence of the warp,
the density $\rho(z,t)=\rho_{eq}(\tilde{x})$ with
$\tilde{x}=z-h_{w}\cos (\nu t+\xi)$ describes an exact solution of the
isothermal Jeans vertical equation for any value of the gas velocity dispersion.
Since this `rigid' mode is reached in a dynamical timescale
$\Omega^{-1}$, a net departure from Jeans hydrostatic equilibrium
because of the warping should be small.
This `incompressible' behaviour has been confirmed
numerically even in the case that the disc evolves
only through its own gravity (see, e.g., \S 4 in Masset 1997).

\subsubsection{Ram pressure and cold gas accretion}
The ram pressure due to
infalling diffuse material from the intergalactic medium
could compress the disc vertically (S\'anchez-Salcedo 2006).
The dynamical collapse time for the H\,{\sc i} layer
is so short that Eq.~(\ref{eq:equilibrium})
is still valid but should be solved with appropriate boundary conditions. 
What we have found is that MOND does not require any hypothetical
and unseen compression by smooth intergalactic accretion flows.

The Galactic disc is probably subject to a continuous process of circulation
of gas and accretion of material through condensing halo gas.
The continuous infall of high velocity clouds onto the H\,{\sc i} disc
may produce enough agitation in the interstellar medium to explain 
the observed levels of turbulence at the outer parts 
(Santill\'an et al.~2007; Booth \& Theuns 2007). 
Even in this case, hydrostatic equilibrium of the thin H\,{\sc i}
layer is still a valid premise, provided that the value of the
turbulent velocity dispersion is taken consistently. 
The scarce impact of massive high velocity clouds may break down
hydrostatic equilibrium but only locally.

\subsubsection{Uncertainties in the surface density of the outer disc}

Our mass model does not include stellar streams as those
detected recently (Newberg et al.~2002;
Ibata et al.~2003; Yanny et al.~2003; Martin et al.~2005;
Mart\'{\i}nez-Delgado et al.~2005).
Since a ring-like mass concentration produces
an enhanced vertical acceleration in MOND, the presence of such structures
are sufficient for explaining the relatively low flaring
as compared to the predicted values at $R\sim 27$ kpc in
Fig.~\ref{fig:happy}. We refrain to find the best-fitting
model until the distribution of the mass in the
stellar disc is measured more accurately.

\subsection{MOND versus CDM}
One is tempted to ask: which competitive model, MOND or dark matter,
reproduces the flaring of our Galaxy with the fewest ad-hoc assumptions? 
Narayan et al.~(2005) and Kalberla et al.~(2007) have explored
the shape and density profile of the dark matter halo using the
flaring technique.
A nearly spherical dark halo as derived from
Sgr dwarf constraints provides a very good fit to the observations 
as long as the dark matter density decays very fast, as $r^{-4}$ 
(Narayan et al.~2005).  In this situation, the total halo mass 
is $2.8\times 10^{11}$ M$_{\odot}$, too low to be acceptable. 
Accretion of homogeneous, diffuse intergalactic gas leads to 
more thin discs (S\'anchez-Salcedo 2006), thereby going in the wrong 
direction to account for the observed flaring. 
A complex dark matter distribution 
(halo + disc + ring) as seen in Kalberla et al.~(2007) may
explain the thickness of the H\,{\sc i} disc but it remains 
unclear if this model is compatible with the orbits of Sgr 
and its stellar debris. Kalberla et al.~(2007) conclude that,
if the dark matter particles only populates a rather spherical component 
with a flattening $0.9\leq q\leq 1.25$ (Helmi 2004; Johnston et al.~2005;
Law et al.~2005; Belokurov et al.~2006; Fellhauer et al.~2006),
then the observed H\,{\sc i} thickness is difficult to
explain unless the effective velocity
dispersion of the gas increases with galactocentric distance.

\begin{figure}
{\rotatebox{270}{\resizebox{11cm}{8cm}{\includegraphics{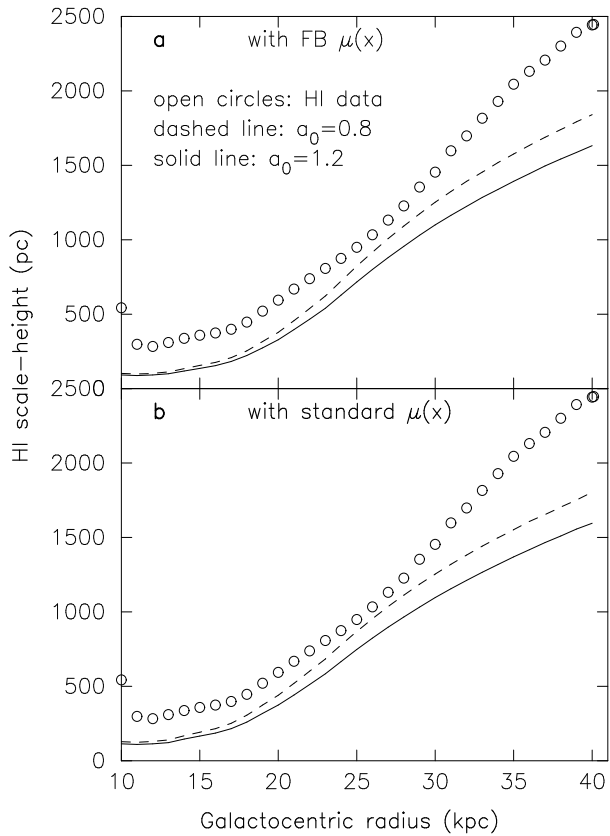}}}}
\caption{H\,{\sc i} scaleheight vs $R$ for the FB (upper panel) and 
the standard (lower panel) interpolating functions and 
$\sigma_{\rm HI}=7$ km s$^{-1}$.  The two curves are generated 
for two values of $a_0$. Lower the value of $a_0$ better is the 
fit to the observed data. This is understandable since lower $a_0$ 
corresponds to lower values of $v_c$ which means less gravitational pull. 
 Beyond $27$ kpc it really deviates from the observation for the 
above model parameters. $a_0$ is in units of $10^{-8}$ cm s$^{-2}$. }

\label{fig:scaleheight}
\end{figure}

\begin{figure}
{\rotatebox{270}{\resizebox{7cm}{8.5cm}{\includegraphics{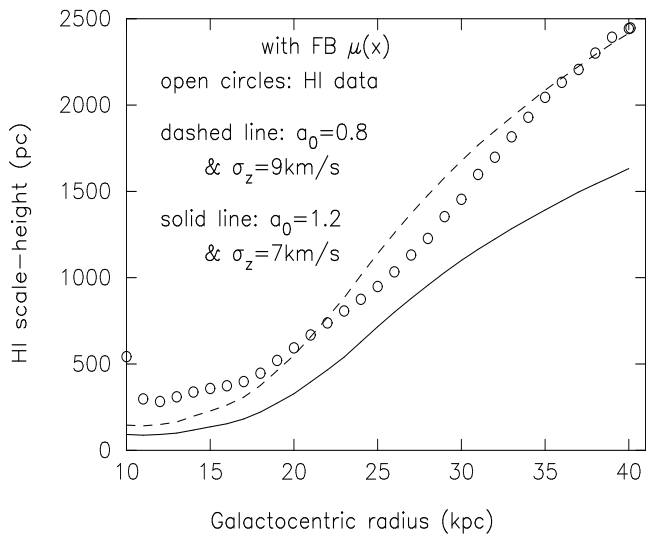}}}}
\caption{H\,{\sc i} scaleheight vs $R$ for 
the FB interpolating function with $a_{0}=0.8\times 10^{-8}$ cm s$^{-2}$  
and $\sigma=9$ km s$^{-1}$. For comparison, the scaleheight
for $a_{0}=1.2\times 10^{-8}$ cm s$^{-2}$ 
and $\sigma=7$ km s$^{-1}$ is also shown.  }
\label{fig:happy}
\end{figure}

\section{Discussion and Conclusions}
Observations of spiral galaxies strongly support a one-to-one analytical
relation between the inferred gravity of dark matter at any radius
and the enclosed baryonic mass. This correlation of gravity with
baryonic mass can be interpreted as a modification of gravity.
If this is not the case, then the success of MOND is telling us
something phenomenological about the nature of dark matter.
Besides the shape of rotation curves of spirals, a wide range of gravitational
effects appear in galactic dynamics, which in
principle, offer independent restrictions on any proposed modified
gravity scheme. Some examples include the dynamics of
stars in the solar neigbourhood (Kuijken \& Gilmore 1987),
the ellipticity of the X-ray
emission in elliptical galaxies (Buote \& Canizares 1994), the
gravitational stability of galactic
discs (e.g., S\'anchez-Salcedo \& Hidalgo-G\'amez 1999), the relaxation
processes in stellar systems (Ciotti \& Binney 2004; S\'anchez-Salcedo
et al.~2006) or the morphology of the Galactic warp (Weinberg \& Blitz 2006). 
In this paper a new attempt to test the self-consistency
of MOND has been carried out. Since in MOND all the gravity comes from
the disc whereas the potential in CDM models is primarily from a
spheroidal distribution, galactic discs are expected to be more
flattened in MOND. For the same reason, if MOND is true, Newtonists
should only infer oblate dark haloes. In this sense, a clear-cut 
inference of a prolate dark halo would be enough to rule out `classical' MOND. 
Interestingly, studies on the shape and flattenings of dark haloes can be 
immediately used as tests to modified gravities.

Our knowledge about the shape of the Galactic dark halo is growing
rapidly. Recent studies of the tidal debris from the Saggitarius
dwarf and its bifurcation constrain the flattening of the dark
halo potential to be close to spherical ($0.9\leq q\leq 1.25$). Contrary to the
naive expectation that MOND should produce far more precession
because all gravity comes from the disc, Read \& Moore (2005)
found the amount of orbital precession is nearly identical to that
which occurs in a CDM Galactic halo with $q=0.9$.

The flaring of H\,{\sc i} have also proven to be useful
to constrain the flattening of dark haloes and is extremely 
sensitive to the dark matter distribution and to the
law of gravity; MOND is expected
to squeeze the disc into a thinner distribution because
of the enhanced vertical acceleration. 
Kuijken \& Gilmore (1989) test MOND in the context of the Oort
discrepancy and find that 
it tends to overcorrect somewhat. However, they used a value of $a_{0}$ 
nearly $4$ times larger than currently measured. Using the updated value
for $a_{0}$, McGaugh \& de Blok (1998) and Famaey \& Binney (2005)
found the dynamics of stars to be consistent with MOND within
the uncertainties.
In this paper we first presented a framework to estimate the 
flaring of H\,{\sc i} layers in MOND.

The MOND field equation is highly non-linear and much more complex
that the classical Poisson field equation. Out of the plane it is
not clear that the algebraic relation (Eq.~\ref{eq:algMOND})
applies.
In addition to the local self-gravity of the disc, MOND predicts a new
term, the $L$-term, which has been shown
to be equivalent to have a fictitious spherical dark matter halo.
The self-gravity term and the $L$-term have very different
radial dependences; the $L$-term is dominant beyond a certain
critical radius. Both terms should work at phase to
explain the observed flaring of galactic discs.
If the $L$-term is ignored, the Galactic flaring is overestimated by
a factor $5$ at the last measured point $R=40$ kpc.

Using this framework, an axisymmetric model was used to derive the mean
H\,{\sc i} flaring of the Milky Way predicted in MOND.
Our Galaxy is particularly interesting because we have information
about some parameters, such as the mass-to-light ratio, that are very difficult
to obtain in other galaxies. 
It is by no means obvious that given the measured
baryonic mass of the Milky Way with essentially no free parameters,
MOND is able to explain the flaring of the H\,{\sc i} disc in
an extended radial range, between $10$ and $40$ kpc.
Note, for instance, that so far the Galactic flaring is only understood
in a dark matter scenario if the Galaxy contains a massive dark disc with
a ring (Kalberla et al.~2007; de Boer et al.~2007).

In order to compare the observed gas-layer flaring to the predictions,
the surface density of the baryonic components (stars and gas)
and the gas velocity dispersion need to be known.  Since the baryonic
distribution is well constrained from observations, the procedure
is essentially with one free parameter $\sigma_{\rm HI}$ between
$7$ km s$^{-1}$ and $9$ km s$^{-1}$ independent of radius.
The results depend on the choice of the Galactic constants,
since these values affect the gas distribution as inferred from
observations. Following the IAU recommendations we take $R_{\odot}=8.5$
kpc and $v_{\odot}=220$ km s$^{-1}$ and we have verified that, given the 
uncertainties, the results are robust to slight changes on the values of  
$R_{\odot}$ and $v_{\odot}$. 

For our reference set of parameters (local
surface density of stars of $45$ M$_{\odot}$ pc$^{-2}$ and $R_{d}=3.2$ kpc), 
we have found that 
MOND successfully reproduces the most recent and accurate
flaring curve beyond $R=17$ kpc
for both standard and FB interpolating functions, provided the
vertical velocity dispersion of H\,{\sc i} is taken $\sim 9$ km s$^{-1}$ 
as that observed in the inner Galaxy (Malhotra 1995; see also
Blitz \& Spergel 1991).
In the interval $10$ kpc $<R<15$ kpc, the observed
scaleheight is $40\%$ and $70\%$ in excess of that predicted by MOND
for the standard ($n=2$) and the FB ($n=1$) interpolating functions,
respectively. 
Clearly, the reduced thickness predicted by MOND is unsatisfactory.
We have verified whether uncertainties in the assumed parameters
can ammilorate this discrepancy 
without altering the good fit at large radii ($R>17$ kpc). 
Interestingly, MOND fails at those radii where the dominant term
in the field equation is not the confining force of the $L$-term
but the local self-gravity component, thus 
$h\propto \mu \sigma^{2}/\Sigma$.
Therefore, a reduction on the adopted local
stellar surface density or, equivalently, on $\Upsilon$
would result in a thicker H\,{\sc i} disc in the region of interest. 
We have seen that by adopting the smallest $\Upsilon$ compatible
with observations, $h$ increases about $27\%$ in the inner disc.
On the other hand,
the scaleheight also depends on the orbital acceleration
$v_{c}^{2}/R$ through $\mu$. A reduction of the adopted value for
$R_{\odot}$ goes in the good direction to mitigate the overcorrection 
but the effect is rather irrelevant. 
For instance, if we assume $R_{\odot}=8$ kpc instead
of $8.5$ kpc, $h$ varies only $\sim 3\%$ at $10$ kpc $<R<15$ kpc.

The simplest way to reconcile predictions with
observations is to argue that ordered and small-scale
magnetic fields and cosmic rays could contribute to support the disc
within the optical disc ($R<15$ kpc). 
Depending on $n$ and $\Upsilon$, the
effective velocity dispersion required to match observations
should be a factor $\sim 1.12$--$1.3$ larger than the turbulent
velocity dispersion. We have seen that, in principle,
magnetic fields and cosmic rays could account for this.
An enhanced effective velocity dispersion within the optical
disc ($R<15$ kpc) is reasonable in a scenario where supernovae 
explosions feed up turbulence.
An enhancement $\sigma_{\rm HI}'/\sigma_{\rm HI}=1.3$
required to explain the flaring when $n=1$ and a local stellar
density of $\Sigma_{\star}=45$ M$_{\odot}$ pc$^{-2}$ is 
likely on the high side. Therefore,  
we can say that MOND satisfactorily explains the flaring 
provided that
$n\geq 1$ and $\Sigma_{\star}<45$ M$_{\odot}$ pc$^{-2}$.

Very little further progress is likely to be made without
further empirical constraints on the effects of non-thermal terms.
If we learn in future that the non-thermal pressure forces from
cosmic rays and magnetic fields have no role to play in the
vertical support of H\,{\sc i} discs (e.g., if the filling factor of
H\,{\sc i} is very low), then studies of the vertical
distribution of cold gas in our Galaxy could either rule out MOND
or, if MOND holds true, pose strong constraints on the stellar 
mass-to-light ratio of the Galactic disc and the form of the 
interpolating function. So far, this analysis is
premature because we believe that a certain level of non-thermal 
support is more realistic.

The maps of H\,{\sc i} surface density and scaleheight in our Galaxy
are far from being axisymmetric. Nevertheless, the narrow anticorrelation
between H\,{\sc i} surface density and the thickness of the H\,{\sc i}
layer (see Figs.~1 and 4 in Levine et al.~2006b) is expected when
the self-gravity of the gas is significant in negotiating the hydrostatic
equilibrium, as occurs naturally in MOND.
Given the present uncertainties in the vertical support by
magnetic fields and cosmic rays, we conclude that 
MOND can plausibly explain the Galactic 
flaring and cannot be excluded as an alternative to dark matter.  
In order to discriminate between 
MOND and dark matter, we plan to extend the analysis for a sample of 
galaxies of different types, including dark dominated LSB and dwarf 
galaxies, with accurate determinations of the rotation curve
and H\,{\sc i} flaring. 

\bigskip

\noindent {\bf{Acknowledgements}}

We are grateful to Peter Kalberla for providing the tables of H\,{\sc i}
surface density and scaleheight from the LAB survey.
We thank the anonymous referee for his/her critical reading of the
manuscript and helpful comments that lead to an improvement of
the paper. We also thank J.~Cant\'o and F.~Masset
for interesting discussions.

\newpage

\appendix
\section{Estimating the magnitudes of $\xi_{\Phi}$ and $\xi_{\mu}$}
\label{sec:vmagnitudes}
In an isothermal self-gravitating disc with one dimensional 
dispersion $\sigma$, the vertical acceleration at a scaleheight 
$h$ is $1.5\sigma^{2}/h$ and the radial acceleration is $v_{c}^2/R$. Thus the ratio between the vertical and radial accelerations is
$\xi_{\Phi}=1.5(\sigma/v_{c})^{2}(R/h)$. In our Galaxy, $\xi_{\Phi}$ 
can be computed for the H\,{\sc i} layer to be $0.14$ and $0.05$ at $15$ 
and $40$ kpc, respectively. 
At regions where $\xi_{\Phi}\ll 1$, i.e. $h>1.5(\sigma/v_{c})^{2}R$,
the radial component determines the kinematics of the disc.
For a galaxy like the Milky Way, this occurs at $R>6$ kpc for the stellar disc and
at $R\gtrsim 1$ kpc for the H\,{\sc i} disc. At these radii, we have
\begin{equation}
|g_{z}| 
\simeq \frac{2\pi G\Sigma(z)}{\mu(x)},
\label{eq:gz}
\end{equation}
and
\begin{equation}
|g_{R}| 
\simeq \frac{v_{c}^{2}}{R},
\end{equation}
with
\begin{equation}
x(R,z)=\frac{|\vec{\nabla}\Phi|}{a_{0}}\simeq \frac{|g_{R}|}{a_{0}}
=\frac{v_{c}^{2}}{Ra_{0}} 
\end{equation}
where $\Sigma(z)\equiv \int_{-z}^{z} \rho (z') dz'$ is the surface
density within a slab of thickness $2z$,
and $v_{c}^{2}(R,z)\equiv R(\partial\Phi/\partial R)$ is the squared planar 
circular velocity.

In order to evaluate $\xi_{\mu}$, we need to estimate $\partial g/\partial z$
at $z\simeq h$.  It is simple to show that
\begin{equation}
\frac{\partial g}{\partial z}\big|_{z=h}\simeq 
2 g_{z}\frac{\partial g_{z}}{\partial z}\big|_{z=h}\simeq
\frac{\pi^{2}G^{2}\Sigma_{\infty}^{2}}{\mu^{2}h}=4\frac{\sigma^{4}}{h^{3}},
\end{equation}
where we have used Eq.~(\ref{eq:gz}) and the relation 
$\sigma^{2}\approx \pi \mu^{-1} G\Sigma_{\infty}h$. 
On the other hand
\begin{equation}
\frac{\partial g}{\partial R}\big|_{z=h}\simeq -2\frac{v_{c}^{4}}{R^{3}}.
\end{equation}

Combining the above equations, we obtain:
\begin{equation}
|\xi(R)| \approx 3\left(\frac{R}{h}\right)^{4}
\left(\frac{\sigma}{v_{c}}\right)^{6}.
\label{eq:xiestimate}
\end{equation}
For illustration, we may compute $\xi$ 
for the Milky Way at $R=15$ kpc and $R=40$ kpc. 
At $R=15$ kpc, $\xi\approx 0.15$ and decreases to $\xi\approx 0.0016$
at $40$ kpc.

\end{document}